\documentclass[letterpaper,10pt]{article}
\usepackage{setspace}
\usepackage{natbib}
\usepackage{amsfonts}
\usepackage{bm}
\usepackage{graphicx}
\usepackage{amsmath,amsfonts,amssymb,amsthm}
\usepackage{xcolor}
\usepackage{float}
\usepackage[T1]{fontenc}
\usepackage[utf8]{inputenc}
\usepackage[english]{babel}
\usepackage{authblk}
\restylefloat{table}
\usepackage[left=2.54cm,top=2.54cm,right=2.54cm,bottom=2.54cm]{geometry}
\title{A higher-order singular value decomposition tensor emulator for spatio-temporal simulators}
\author[1]{Giri Gopalan}
\author[2]{Christopher K. Wikle}
\begin{document}
\affil[1]{Statistics Department, California Polytechnic State University, San Luis Obispo, California, 93407 (ggopalan@calpoly.edu)}
\affil[2]{Department of Statistics, University of Missouri, Columbia, Missouri, 65211 (wiklec@missouri.edu)}

\maketitle
\begin{abstract}
\singlespacing
We introduce methodology to construct an emulator for environmental and ecological spatio-temporal processes that uses the higher order singular value decomposition (HOSVD) as an extension of singular value decomposition (SVD) approaches to emulation. Some important advantages of the method are that it allows for the use of a combination of supervised learning methods (e.g., random forests, and Gaussian process regression) and also allows for the prediction of process values at spatial locations and time points that were not used in the training sample. The method is demonstrated with two applications: the first is a periodic solution to a shallow ice approximation partial differential equation from glaciology, and second is an agent-based model of collective animal movement. In both cases, we demonstrate the value of combining different machine learning models for accurate emulation. In addition, in the agent-based model case we demonstrate the ability of the tensor emulator to successfully capture individual behavior in space and time.  We demonstrate via a real data example the ability to perform Bayesian inference in order to learn parameters governing collective animal behavior. 
\end{abstract}
\section*{Acknowledgements}
This work was partially supported by the U.S. National Science Foundation (NSF) through NSF Award DMS-1811745.

\newpage
\doublespacing
\section{Introduction}

Multivariate spatio-temporal processes are ubiquitous in agricultural, biological, and environmental science. Modeling such processes is complicated by complex dependencies across processes, time, and space, as well as uncertainty in observations, process specification, and parameters that are contained in the observation and process models.  Since the mid-1990's, Bayesian hierarchical models have been the primary tool to consider such processes in statistics \citep{10.1007/978-94-011-5430-7_3}. Although there are challenges to each component of such a model, for multivariate spatio-temporal processes, arguably the greatest challenge is the specification of a model for the process that is rich enough, yet parsimonious enough, to capture realistic behavior (e.g., nonlinear interactions, non-stationarity, non-separability, collectiveness, etc.). One approach to this problem has been to specify mechanistically-motivated dynamical parameterizations within the hierarchical statistical framework (i.e., "physical-statistical models", \cite{Berliner,kuhnert2014physical}) such that they include the potential for known mechanistic behavior (e.g., diffusion, advection, density dependent growth, etc.; see \citet{wikle2010general}).  In situations where one has a greater knowledge of the fundamental process dynamics, it is often preferable to embed a "black box" dynamical model (simulator) within a formal statistical framework, such as is done in the computer model calibration literature \citep{kennedy2001bayesian}. 
Examples of such black box simulators include partial differential equation (PDE) models  \citep{gopalan2019hierarchical} or agent-based models (ABMs) that have multiple individuals interacting across time and space \citep[e.g.,][]{fadikar2018calibrating}.

The intersection of these approaches for the statistical modeling of complex dynamical processes occurs when the black-box models are too expensive to implement within statistical inference procedures (typically Bayesian), and one must build an emulator (i.e., surrogate statistical model) to represent the input-output relationships in the black-box model. Essentially, the emulator is a function that mimics the output of the simulator but is computationally less expensive to evaluate. Emulators are most often specified by Gaussian processes (GPs), but non-linear surrogate models can also be used (see the overview in \citet{gramacy2020surrogates}).  

Although surrogate modeling in statistics is a mature and well-studied topic, there are still challenges with respect to developing efficient surrogates for black-box models with high-dimensional multivariate spatio-temporal output. Perhaps the most common approach to dealing with spatio-temporal model output in the computer model calibration literature has been to project the model output on a reduced-rank set of basis functions as in principal components analysis (PCA or singular value decomposition, SVD) \citep{higdon2008computer}.  In this case, one typically models the lower-dimensional PC coefficients in terms of GPs.  However, different basis functions can be used \citep{doi:10.1080/01621459.2018.1514306} and one can also model the relationship between the inputs and projection coefficients (right singular vectors) via non-linear functions in the mean. This is important because often the relationship between the inputs and the right singular vectors is not smooth. 

It is much less common to build surrogate models for multivariate or multi-state spatio-temporal data. \citet{leeds2014emulator} considered a surrogate model for a multivariate spatio-temporal process consisting of phytoplankton, sea surface temperature, and sea level pressure. A more recent exception is \citet{pratola2018bayesian}, who made use of the so-called "higher order singular value decomposition" (HOSVD) tensor analog of SVD \citep{de2000multilinear} to build a GP surrogate model for multi-state spatio-temporal data. However, it is important to realize that input-output structure can be quite different whether one considers space, time, multiple-processes, or parameters.  As mentioned above, it is quite possible that such relationships are not smooth and that in some cases it might be better to model the relationships in the mean (first-order) and in other cases in the covariance (second-order), as via a GP.  Thus, we extend the first-order SVD modeling approach to the HOSVD tensor framework. In essence, the proposed tensor-based method stores a tensor of simulator runs and performs a HOSVD; then, machine learning models are trained to learn the analogs of right singular vectors.  Importantly, we are free to select different sub-emulators for space, time, processes, and parameters in this framework. 

Our goal is to demonstrate this HOSVD approach for generating emulators for complex input-output structures. Such an emulator can then be used either as a pure multivariate spatio-temporal prediction model that uses only simulator runs for training, for model calibration given field data, or to perform inference on model parameters, depending on the goals of the researcher.  Thus, the contributions of this methodology are:
\begin{enumerate}
\item It provides for a combination of different supervised learning approaches (e.g., random forests (RFs) and GP regression) that can be used to emulate the spatial, temporal, and parameter components of the model output as obtained through the HOSVD. 
\item Parameters can be batched together in the tensor-decomposition, which is useful if parameters are correlated \textit{a priori}.
\item Surrogate model predictions can be made at locations different from the set of spatial and temporal locations used in the training runs (in contrast to many SVD approaches for emulating high-dimensional output).
\item Multiple individuals in an ABM with spatio-temporal locations can be emulated efficiently in space and time.
\item Computational savings are afforded by using a reduced rank tensor factorization through the truncated HOSVD \citep{de2000multilinear}; a low-rank tensor factorization is usually sufficient to capture much of the variation in simulator output, just as a few principal components are usually sufficient to capture most of the variation in lower-dimensional simulator output.
\end{enumerate}

The remainder of the paper is organized as follows. In Section 2, we briefly review aspects of SVD emulation, and tensors. In Section 3, we introduce a statistical model in terms of the HOSVD and describe the method for emulation based on this model. In Section 4, we provide empirical results of the application of the tensor-based method in two scenarios: i) a periodic solution to the shallow ice approximation from glaciology \citep{Bueler}, which is used to demonstrate the advantage of predicting at locations not considered in the training, and ii) a simulation of an ABM for collective animal movement to illustrate that the emulator can successfully capture the locations of multiple individuals moving collectively in space and time. These examples demonstrate the tensor method's ability to apply to predictions off of a grid, flexibility in combining different unsupervised learning methods, and use in Bayesian inference. We present a brief discussion and conclusion in Section 5.

\section{SVD Emulation and Tensor Background}
Before introducing the statistical model and associated methodology of the HOSVD tensor-based emulator approach that is the subject of the paper, we provide a brief introduction to emulators, a review of the approach from \citet{higdon2008computer} and \cite{Hooten2011}, and an overview of tensors, as our approach builds on these methods.  

\subsection{Emulators/Surrogates}
The use of emulators, or equivalently surrogate models, has appeared widely in the uncertainty quantification and computer experiments literature, as is detailed in \cite{gramacy2020surrogates}. Their principal use is to approximate computationally intensive functions at a reduced computational cost. Very broadly, they can be categorized as methods that use GPs, as in \cite{gu2018} for instance, and polynomial basis (polynomial chaos) expansions, as in \cite{Sargsyan2016}.  In addition, others have considered surrogate models that are based on mechanistic models \citep[e.g.,][]{leeds2014emulator} and neural network models \citep[e.g.,][]{tripathy2018deep}.  This is an active area of research.  For example, \cite{kasim2020up} recently introduced a method that uses a deep neural network that learns its own architecture for the purposes of emulation.

Below, we discuss in more detail emulators that are based on modeling the right singular vectors in an SVD and come back to tensor-based emulators after a brief introduction to tensor decomposition. 

\subsection{SVD Emulators}
The proposed tensor-based emulator is an extension of SVD-based emulators.  Here we describe two approaches to SVD emulation - a GP (second-order) approach in \cite{higdon2008computer} and a random-forest (first-order) approach in \cite{Hooten2011}. 
\cite{higdon2008computer} construct Bayesian hierarchical models involving observational data in the manner of \cite{kennedy2001bayesian} in order to calibrate physically important parameters and to make predictions with uncertainties. Because the physics-based computer simulators they employ are computationally expensive to evaluate, they use a basis of principal components of simulator runs (derived with the SVD), and use GP prior distributions on the weights associated with the linear combination of basis vectors derived from the SVD. \cite{Hooten2011} utilize a similar SVD approach for constructing an emulator, except they relax the constraint that the weights in the linear combination have GP prior distributions, and consider instead other functional forms in the mean response, such as random forests. An extension of this approach to the tensor case is one of the features of our methodology.

First, assume that computer simulator runs have been placed in a matrix, $\boldsymbol{C}$, where each column of $\boldsymbol{C}$ corresponds to a distinct computer simulator run. If there are $N$ computer simulator runs for distinct parameter values $\boldsymbol{\theta}$, with simulator output of dimension $M$, then the matrix $\boldsymbol{C}$ is of dimension $M$ by $N$. The first-order emulator relies on using the singular value decomposition, $\boldsymbol{C} = \boldsymbol{U}\boldsymbol{D}\boldsymbol{V}^{\textrm{T}}$, where $\boldsymbol{U} \in \mathbb{R}^{M \times M}$ and $\boldsymbol{V} \in \mathbb{R}^{N \times N}$ are orthonormal and $\boldsymbol{D} \in \mathbb{R}^{M\times N}$ is diagonal. In particular, a statistical model for the $M$-dimensional computer simulator output, $\boldsymbol{c}$, given a parameter,  $\boldsymbol{\theta}$, is:
\begin{eqnarray*}
\boldsymbol{c} = \boldsymbol{U}\boldsymbol{D}\boldsymbol{v}(\boldsymbol{\theta}) + \boldsymbol{\epsilon}
\end{eqnarray*}
for a residual term $\boldsymbol{\epsilon}$ that has mean 0 and possibly non-diagonal covariance matrix $\boldsymbol{\Sigma}$. A multivariate normal random variable for $\boldsymbol{\epsilon}$ is suggested, though is not necessary. 

The central notion encapsulated by the above model is that a vector of computer simulator output can be approximately expressed as a linear combination of the columns of $\boldsymbol{U}\boldsymbol{D}$, and the function $\boldsymbol{v}(\boldsymbol{\theta})$ specifies the $N$ coefficients of the linear combination for a particular value of $\boldsymbol{\theta}$; as such, the function $\boldsymbol{v}(\boldsymbol{\theta})$ has $N$-dimensional vector output. The approach of \cite{Hooten2011} is to model $\boldsymbol{v}(\boldsymbol{\theta})$ as $\boldsymbol{g}(\boldsymbol{\theta}, \boldsymbol{\beta})$, where $\boldsymbol{g}$ is a non-linear regression function that involves $\boldsymbol{\beta}$ as tuning parameters and takes as input the parameter of inferential interest, $\boldsymbol{\theta}$. Each component of the vector $\boldsymbol{g}(\boldsymbol{\theta}, \boldsymbol{\beta})$ uses a non-linear regression model, as for instance with a random forest: the training data come from the $N$ input values of $\boldsymbol{\theta}$ for the computer simulator runs, and the outputs are the rows of $\boldsymbol{V}^{\textrm{T}}$. In contrast, \cite{higdon2008computer} uses a prior distribution with 0 mean GPs in order to learn the coefficients for the singular vectors along with other model parameters.

Computational cost reduction stems from using less than $N$ columns of $\boldsymbol{U}\boldsymbol{D}$, in the way that only a few principal components are used in a principal components analysis. The first few columns of $\boldsymbol{U}\boldsymbol{D}$ are usually sufficient to capture most of the variation of the simulator output (in previous examples, $r$ = 3 was sufficient to capture more than 99\% of the variation). This leads to a reduced computational cost for a few reasons: if the first $r$ columns ($< N$) of $\boldsymbol{U}\boldsymbol{D}$ are used, then $\boldsymbol{v}(\boldsymbol{\theta})$ is $r$ dimensional, so only $r$ machine learning models are needed for $\boldsymbol{v}(\boldsymbol{\theta})$, reducing the time needed to evaluate $\boldsymbol{v}(\boldsymbol{\theta})$ and compute $\boldsymbol{U}\boldsymbol{D}\boldsymbol{v}(\boldsymbol{\theta})$.  

\subsection{Tensor Background}
A real-valued tensor is a multidimensional array of real numbers, just as a matrix is a real-valued two-dimensional array of real numbers. In this subsection, our intention is to provide a brief review of aspects of tensors that will enable the reader to understand the tensor emulator that we utilize in Section 3. A comprehensive reference for tensors is \cite{kolda2009tensor}.

We focus on the HOSVD since it is arguably the most natural extension of the SVD.  Assume that $\boldsymbol{X}$ is a $K$-tensor with dimensions $n_1$, $n_2$, ..., $n_K$. The higher order singular value decomposition of $\boldsymbol{X}$, or HOSVD($\boldsymbol{X}$), decomposes $\boldsymbol{X}$ as $\boldsymbol{Z} \times \boldsymbol{U}_1 \times \boldsymbol{U}_2 \times ... \times \boldsymbol{U}_K$. The properties of the decomposition are:

\begin{itemize}
\item $\boldsymbol{Z}$, referred to as the \textit{core tensor}, is a completely orthogonal tensor whose dimensionality is the same as the dimensionality of $\boldsymbol{X}$. Being completely orthogonal means that any two sub-tensors along the same coordinate have a 0 inner-product; the inner-product between real-valued tensors multiplies coordinates of the same type and then sums the resultant products.

\item $\boldsymbol{U}_1,\ldots,\boldsymbol{U}_K$ are matrices with dimensions $n_1$ by $n_1$, $n_2$ by $n_2$, ..., and $n_K$ by $n_K$, respectively. Additionally, $\boldsymbol{U}_1,\ldots,\boldsymbol{U}_K$ are orthogonal, analogous to the usual SVD.
\end{itemize}

In order to interpret the tensor decomposition discussed previously, one must understand tensor-matrix multiplication, which is described in terms of the familiar matrix-vector multiplication operation. If $\boldsymbol{Z}$ can be thought of as $n_2\times n_3 \times ... \times n_K$ length $n_1$ column vectors; then $\boldsymbol{Z} \times \boldsymbol{U}_1$ is an $n_1$ by $n_2$ ... by $n_K$ tensor where each column vector of length $n_1$ in $\boldsymbol{Z}$, denoted as $\boldsymbol{z} \in \mathbb{R}^{n_1\times1}$, is replaced with $\boldsymbol{U}_{1}\boldsymbol{z}$. Successive multiplication with $\boldsymbol{U}_2,..., \boldsymbol{U}_K$ proceeds in the same manner. 

The specific use of tensors for surrogate modeling appears in the statistics literature in \cite{pratola2018bayesian}. Building off of this work, the method presented in Section \ref{sec:tensoremulator} allows for spatio-temporal predictions that are off of the ``grid'' of observation locations and also allows for the flexibility to use a combination of different supervised learning methods, in addition to GPs. Additionally, a regularized regression approach for a low-rank tensor approximation appears in \cite{chevreuil2015least} and a low-rank polynomial tensor basis expansion appears in \cite{KONAKLI20161144}; in contrast, the methodology presented here uses the truncated HOSVD \citep{de2000multilinear} with a combination of existing supervised learning methodology (i.e., GP regression and RFs).

\section{HOSVD Tensor Emulation}\label{sec:tensoremulator}
In this section, we develop a statistical emulator by using the HOSVD. An important advantage of this approach is being able to emulate each component of a multi-dimensional spatio-temporal input-output space as well as model parameters by including separate surrogate models in each dimension. Additionally, an important aspect is the selection of different machine-learning models for the different tensor components, which is explained in Section 3.3.  

\subsection{Model Formulation}
Without loss of generality, assume we seek to emulate a nonlinear function $f(x,y,t,\theta_1,..., \theta_p)$, where $(x,y,t)$ are spatio-temporal coordinates and $(\theta_1,..., \theta_p)$ are parameters, compactly referred to as $\boldsymbol{\theta}$. The parameters can be either vector or scalar quantities. Assume one has evaluated $f$ for all combinations of $M$ $x$-coordinates, $N$ $y$-coordinates, $T$ $t$-coordinates, $P_1$ values of $\theta_1$, $P_2$ values of $\theta_2$,  ..., $P_p$ values of $\theta_p$. (So $M\times N\times... \times P_p$ evaluations of the function $f$.) These evaluations (e.g., computer simulator runs) can be stored in an $M$ by $N$ by $T$ by $P_1$ ... by $P_p$ dimensional tensor, $\boldsymbol{X}$, where the $i,j,k,l_1,..., l_p$ cell of the tensor stores the function evaluation under the $i$-th $x$ coordinate, $j$-th $y$ coordinate, $k$-th $t$ coordinate, $l_1$-th parameter value of $\theta_1$,..., to $l_p$-th parameter value of $\theta_p$. 

It is important to note that there are multiple variations possible for the initial function evaluations of $f$. For example, the $x$-coordinates, $y$-coordinates, and $t$-coordinates can be considered jointly (e.g., as $(x,y)$ tuples or $(x,y,t)$ tuples). Furthermore, an additional spatial dimension can be considered (i.e., $(x,y,z)$) if desired. Additionally, scalar parameters $\theta_i$ and $\theta_j$ can be considered jointly as $(\theta_i,\theta_j)$ tuples. Such an approach is illustrated in the first example of Section 4, where $(x,y)$ tuples and amplitude and period parameters are sampled jointly. The initial values chosen to populate $\boldsymbol{X}$ may vary from application to application; for instance, a pre-defined lattice of $(x,y)$ coordinates may be used in the context of a numerical PDE solver. In the absence of prior distributions from which to sample parameters, we suggest using a latin hypercube design or some hybrid variant due to its prevalence in the computer experiments literature \citep{gramacy2020surrogates}.

Another important variation is a scenario where one is emulating a spatio-temporal ABM, where, without loss of generality, the function $f$ has both $x$ positions ($f_x$) and $y$ positions ($f_y$). In that scenario, $f_x$ and $f_y$ take the form $f_x(i,t,\theta_1,..., \theta_p)$ and $f_y(i,t,\theta_1,..., \theta_p)$, where $i$ indexes over the agents, $t$ stands for time, and $\theta_1,..., \theta_p$ represent the parameters. Tensor emulators are then constructed separately for \textit{both} $f_x$ (the $x$ values of the trajectories) and $f_y$ (the $y$ values of the trajectories). An example of this variation is illustrated in the second application. The tensor-based methodology proceeds in the same essential manner as follows.

Let $\boldsymbol{Z} \times \boldsymbol{U}_1 \times \boldsymbol{U}_2 \times ... \times \boldsymbol{U}_{p+3}$ be the HOSVD of $\boldsymbol{X}$. The statistical model for an  evaluation of $f$ at $(x*,y*,t*,\theta*_1, ..., \theta*_p)$ is:
\begin{eqnarray*}
f(x*,y*,t*,\theta*_1, ..., \theta*_p) &=& \boldsymbol{Z} \times \boldsymbol{u}_1(x*) \times \boldsymbol{u}_2(y*) \times \boldsymbol{u}_3(t*) \times ... \times \boldsymbol{u}_{p+3}(\theta*_p) + \epsilon,
\end{eqnarray*}
for a residual term $\epsilon$ where $E[\epsilon] = 0$ and $\boldsymbol{u}_1,..., \boldsymbol{u}_{p+3}$ are nonlinear, vector-valued functions. That is, the functions $\boldsymbol{u}_1$ through $\boldsymbol{u}_{p+3}$ behave like the function $\boldsymbol{v}$ from the SVD approach described in Section 2.2. In particular, $\boldsymbol{u}_1(.): \mathbb{R} \rightarrow \mathbb{R}^{1\times M}$, $\boldsymbol{u}_2(.): \mathbb{R} \rightarrow \mathbb{R}^{1\times N}$, and so on. The most basic model assumes that the error term $\epsilon$ is independent and identically Gaussian, though that is not a necessary requirement. For instance, \cite{gopalan2019hierarchical} consider a multivariate random walk with spatio-temporal correlation. 

\subsection{Model Implementation}

Our purpose in this section is to describe in detail how to construct an emulator (i.e., how to obtain an estimator $\hat{f}$) using the HOSVD components of the tensor simulator runs.

\subsection{Emulator Construction}\label{sec:emulatorconstruction}
Assume that $\boldsymbol{X}$ is a tensor of function evaluations as in the previous section, with HOSVD that is given by $\boldsymbol{Z} \times \boldsymbol{U}_1 \times \boldsymbol{U}_2 \times \boldsymbol{U}_3 \times ... \times \boldsymbol{U}_{p+3}$ for core tensor $\boldsymbol{Z}$ and matrices $\boldsymbol{U}_1$ through $\boldsymbol{U}_{p+3}$. The methodology presented is to develop estimators for functions $\boldsymbol{u}_1,..., \boldsymbol{u}_{p+3}$; these are denoted as $\boldsymbol{\hat{u}}_1, ..., \boldsymbol{\hat{u}}_{p+3}$. Then, the estimated value for $f(x*,y*,t*,\theta*_1, ..., \theta*_p)$ is
\begin{eqnarray*}
\hat{f}(x*,y*,t*,\theta*_1, ..., \theta*_p) &=& \boldsymbol{Z} \times \boldsymbol{\hat{u}}_1(x*) \times \boldsymbol{\hat{u}}_2(y*) \times \boldsymbol{\hat{u}}_3(t*) \times ... \times \boldsymbol{\hat{u}}_{p+3}(\theta*_p).
\end{eqnarray*}
First, consider the function $\boldsymbol{u}_1(x)$, which when expanded as a vector is $(u_{11}(x), ..., u_{1M}(x))$, an element of $\mathbb{R}^{1\times M}$. Specifically, $\hat{u}_{11}(x)$ is obtained by using a standard supervised learning approach, such as GP regression or a random forest. The training data for learning $u_{11}$ are the $M$ values of $x$ (i.e., the features) and the first column of $\boldsymbol{U}_1$ (i.e., the responses). The training data for learning $u_{12}$ are the $M$ values of $x$ and the second column of $\boldsymbol{U}_1$ and so on, up to $u_{1M}$, which uses the $M$th column of $\boldsymbol{U}_1$. The same estimation procedure is applied for $\boldsymbol{\hat{u}}_1, \boldsymbol{\hat{u}}_2,...$, and $\boldsymbol{\hat{u}}_{p+3}$.

An important aspect of the above procedure is the selection of the machine learning models for learning the $\boldsymbol{u}$ functions. While in practice this must be achieved with experimentation in a real-data scenario (as in the application examples that follow), one reason why this is potentially advantageous is because not all components of the $\boldsymbol{u}$ functions are going to be very smooth (i.e., continuously differentiable), so considering alternate machine learning models besides a GP regression is sensible. (An example is presented in the supplementary figures.)

The package \textit{rTensor} \citep{JSSv087i10} provides an option for a reduced rank tensor factorization through the truncated HOSVD \citep{de2000multilinear}. A reduced rank factorization does not completely recover the input tensor, $\boldsymbol{X}$, but it approximates $\boldsymbol{X}$ as $\boldsymbol{Z} \times \boldsymbol{V}_1 \times \boldsymbol{V}_2 \times \boldsymbol{V}_3 \times ... \times \boldsymbol{V}_{p+3}$ where the number of columns (i.e., \textit{ranks}) of $\boldsymbol{V}_1$, $\boldsymbol{V}_2$,..., $\boldsymbol{V}_{p+3}$ are no more than $M$, $N$, $T$, ...,  $P_p$, respectively. The number of rows, however, remain to be $M$, $N$, $T$, ...,  $P_p$ of $\boldsymbol{V}_1$, $\boldsymbol{V}_2$, $\boldsymbol{V}_3$, ...,  $\boldsymbol{V}_{p+3}$, respectively. The core tensor $\boldsymbol{Z}$ has dimensions equal to the column numbers of $\boldsymbol{V}_1$, $\boldsymbol{V}_2$,..., $\boldsymbol{V}_{p+3}$, respectively. Using the low-rank approximation is computationally advantageous for the same reason that one uses only the top few of the columns of $\boldsymbol{U}\boldsymbol{D}$ for emulation in the first-order SVD case; the number of machine learning models needed is reduced based on the rank, which in turn speeds up the emulator evaluation and tensor-matrix multiplication.

\subsection{Selection of Ranks}
The starting point of the procedure outlined above is to take the HOSVD decomposition of the tensor of simulator runs ($\boldsymbol{X}$), for which it may be computationally beneficial to use a low-rank representation via the truncated HOSVD. Here we provide guidance for selecting the ranks for the truncated HOSVD in a principled manner, if the user chooses to use a low-rank approximation. The method iteratively performs rank selection via SVD scree plots for each mode of the tensor after tensor unfolding. Specifically, the suggested method is as follows: 
\begin{enumerate} 
\item For each mode of the tensor $\boldsymbol{X}$, unfold the tensor. For example, if the tensor has dimensions $n_1$, $n_2$, $n_3$, and $n_4$, unfolding the tensor by the first mode yields a matrix with $n_1$ rows and $n_2n_3n_4$ columns, where row $i$ groups all of the tensor elements where the first mode is $i$ (of $n_1$). The tensor unfolding operation, along with a computational implementation, is described in \citet{JSSv087i10}.
\item For each mode of $\boldsymbol{X}$, take the SVD of the unfolded and centered tensor and use a standard procedure to determine the total number of singular vectors to keep. (As is usual, centering refers to ensuring column sample means are 0.) Specifically, we use a scree plot of singular values along with a comparison to singular values of a randomized matrix (each column shuffled) to determine a singular value threshold for calculating the total number of singular vectors to keep, as suggested in Chapter 14 of \citet{hastie2009elements} for PCA.
\item For each mode, set the rank in the truncated HOSVD to the number of retained singular vectors from the previous step, except adding 1 to account for the mean that is subtracted when centering the data.
\end{enumerate}
To check the quality of the low rank tensor decomposition that is found with the previous steps, we suggest calculating the proportion of variance explained by the low rank decomposition with the familiar formula:
\begin{eqnarray*}
1- RSS/TSS,
\end{eqnarray*}
where RSS stands for residual sum of squares and TSS stands for total sum of squares. Here, RSS is the square of the Frobenius norm of the residual tensor, and TSS is the squared Frobenius norm of the training tensor minus the grand mean. That is -- let $\boldsymbol{X}$ be the training tensor as usual, and let $\boldsymbol{R}$ be the low rank approximation via the truncated HOSVD. Finally, let $\boldsymbol{X}_c$ be the tensor formed by subtracting every element of $\boldsymbol{X}$ by $\bar{X}$, a scalar that is the mean of all elements of $\boldsymbol{X}$ (i.e., the grand mean). Then 
\begin{eqnarray*}
RSS &=& {||\boldsymbol{X}-\boldsymbol{R}||_F}^2, \\
TSS &=& {||\boldsymbol{X}_c||_F}^2,
\end{eqnarray*}
where the subscript $F$ denotes the Frobenius norm. The proportion of variance explained by the low-rank approximation is more interpretable than the Frobenius norm of the residual.

\section{Examples}
We consider two examples to demonstrate some of the advantages of the HOSVD tensor emulator. Specifically, in Section \ref{sub:glacier} we consider an emulator for a periodic exact solution to the shallow ice approximation partial differential equation (SIA PDE) from \citet{Bueler}, and in Section \ref{sub:animal} we consider an emulator for an ABM simulation of collective animal movement. Both examples show the advantages of considering different surrogates for spatial components and parameters. In addition, the SIA PDE example demonstrates how our emulator approach can generate spatial predictions at spatial locations and time points that were not observed, and the collective movement ABM example demonstrates that our approach can reasonably emulate the locations in space and time of multiple individuals.  Both experiments make use of the \textit{rTensor} \citep{JSSv087i10} R package to find the HOSVD and to perform tensor matrix multiplication. 

\subsection{Oscillating Glacier Example}\label{sub:glacier}
The SIA PDE is a commonly used mathematical model that describes the time evolution of glacier thickness in glaciers that are shallow; e.g., see \cite{Tolly, flowers2005sensitivity, alex, 2016arXiv161201454G, werder_huss_paul_dehecq_farinotti_2020}. The main physical principle utilized in this model is mass conservation, and the shallowness approximation allows one to ignore stress terms in formulating the PDEs. 

Here we use a HOSVD tensor-based emulator for a periodic solution to the SIA PDE described in \cite{Bueler}. The periodic solution is a dome that oscillates according to a particular period and amplitude. This periodic function is used in \cite{gopalan2019hierarchical}, and is repeated here for completeness:
\begin{eqnarray*}
H(r,t) &=& H_s(r)+P(r,t), \\
P(r,t)  &=& C_p\sin(2\pi t/T_p)\cos^2\left[\frac{\pi(r-0.6L)}{.6L}\right]; \textrm{if } 0.3L < r < .9L, \\
P(r,t) &=& 0; \textrm{if } 0 \leq r \leq 0.3L \textrm{ or if } r \geq 0.9L. 
\end{eqnarray*}
In this formulation, $H$ stands for glacier thickness, and is a function of distance from the origin, $r$, and time, $t$; in particular, this is a radially symmetric solution. The term $H_s$ (Eq. 21 of \cite{Bueler}) corresponds to a static profile that is added to a time-varying periodic function, $P$, that has an amplitude $C_p$ and period $T_p$. Here, $L$ refers to the length of the glacier profile. Overall, the solution looks like a dome with a periodic oscillation in thickness within the domain $0.3L < r < .9L$. The parameters that we focus on for this example are $C_p$, the amplitude of the periodic perturbation, and $T_p$, the period of the perturbation. For more details of this model and the periodic solution, see \cite{Bueler}. 


\subsubsection{Glacier Example Experimental Design}
To demonstrate the HOSVD tensor emulator for the glacier dynamics problem we consider the following experiment. For $s = 10, 20, 30$:
\begin{enumerate}
    \item $s^2$ 2-tuples of $(x,y)$ are drawn using latin hypercube sampling on the range $-5\times 10^5 m$ to $5\times 10^5 m$ for the $x$ and $y$ location values.
    \item $s$ values of $t$ are sampled uniformly from 0 to 10000 years. 
    \item $s^2$ 2-tuples of (period, amplitude) are drawn using latin hypercube sampling on the range (1000 to 5000) years for period and (100 to 400) meters for amplitude. 
    \item For each combination $(x,y)$, (period, amplitude), and $t$, the glacier thickness function that is to be emulated is evaluated and the result is stored in a tensor, $\boldsymbol{X}$. The dimensions of the tensor are $s^2$ by $s$ by $s^2$, so there are $s^5$ total elements in the tensor.
    \item The methodology from Section 3 is applied to the tensor $\boldsymbol{X}$. The combinations of machine learning models tried are all random forests for the spatial, temporal, and parameter components, all GPs and RFs for the parameter component, but GPs for the spatial and temporal components. We implemented these machine learning models with the \textit{kernlab} R package \citep{Karatzoglou:2004aa}) and the \textit{randomForest} R package \citep{randomForest}. 
    \item The predictive accuracy of the emulator is assessed by calculating the mean absolute relative error on a test set. Absolute relative error is $|(H_{true} - H_{pred})/H_{true}|$. The test set is randomly selected with the same design steps as above, and a test set of size 100 is used; because the test $(x,y)$ and $t$ values are not necessarily the same as the training $(x,y)$ and $t$ values, this example illustrates the capability of the tensor-based method to emulate locations and times that were not in the training sample.
\end{enumerate}

Note that since $(x,y)$ and (period, amplitude) are considered jointly, we end up with only three $\boldsymbol{U}$ matrices, $\boldsymbol{U}_1, \boldsymbol{U}_2,\boldsymbol{U}_3$, where $\boldsymbol{U}_1$ corresponds to $(x,y)$, $\boldsymbol{U}_2$ corresponds to $t$, and $\boldsymbol{U}_3$ corresponds to (period, amplitude). Additionally, we used the truncated HOSVD from \textit{rTensor} and the rank selection procedure detailed in Section 3.4. The supplementary figures show the scree plots used for determining ranks, and the proportion of variance explained by the low rank approximation is .99. 

\subsubsection{Glacier Example Results}
The results of the previously described experiment are illustrated in Figure 1. The $x$-axis of this plot represents a comparison of the mean absolute relative error over all sizes, which indicates that the absolute relative error decreases as the size of the training tensor increases.  In addition, this plot shows that the HOSVD approach that considers mixed machine learning models yields the smallest error. Specifically, the emulator that included a GP surrogate for $(x,y)$, GP surrogate for $t$, and random forest surrogate for (period, amplitude) performed better than pure random forest or GP approaches in terms of error. Additionally, we repeated the experimental design ten times in order to assess the variability in the predictions with respect to the initial input $(x,y)$, $t$, and (period, amplitude) values. Boxplots comparing the absolute relative errors are shown in Figure 2, again showing that for this example, the mixture of machine learning surrogates tends to perform best.  In addition, a comparison of the actual and emulated values of glacial thickness using the mixture of machine learning models tensor emulator is shown in Figure 3.  This shows that there is generally a good agreement between the actual thickness and predicted thickness across thickness values, with smallest errors towards the higher glacier thickness values; one possible explanation for this behavior is that the magnitude of periodic movement for this particular example is largest within the interior of the glacier, at lower thickness values. 

\begin{figure}[h!]
  \centering
    \includegraphics[width=0.8\textwidth]{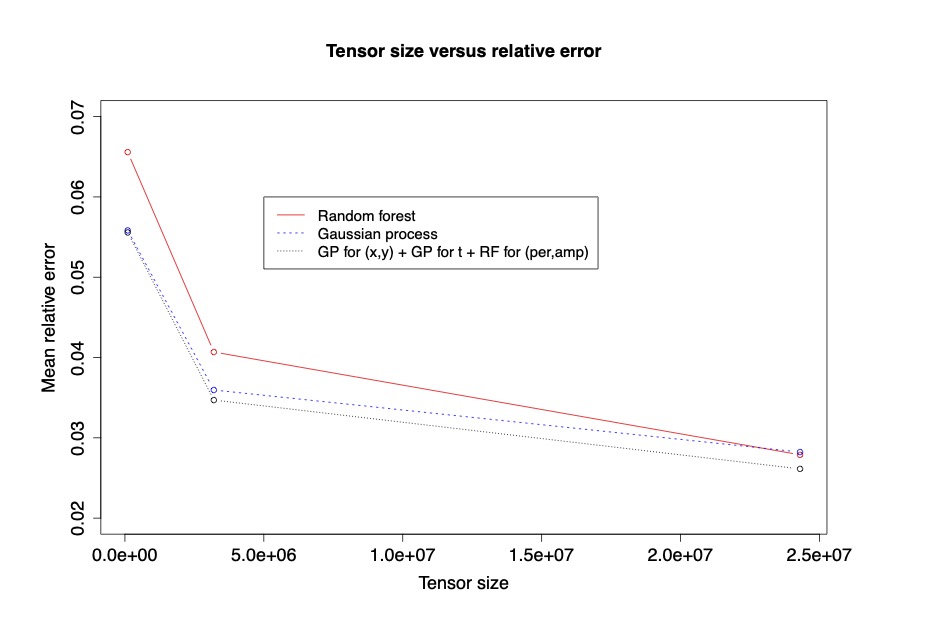}
\caption{Comparison of tensor size and mean absolute relative error, for $s = 10, 20,$ and $30$ (see Section 4.1.1) -- generally, the error decreases with a larger tensor and also is smallest with a combination of machine learning models.}
\end{figure}

\begin{figure}[h!]
  \centering
    \includegraphics[width=0.8\textwidth]{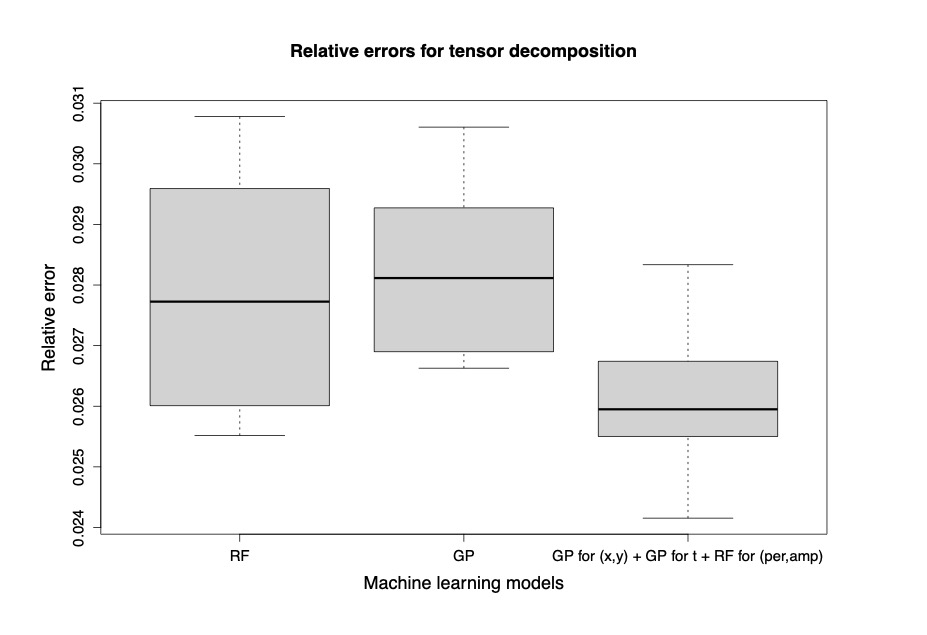}
\caption{Comparison of mean absolute relative error for a pure random forest, pure GP, and a GP for $(x,y,t)$ and random forest for (period, amplitude). The mixture of machine learning models generally performs the best.}
\end{figure}

\begin{figure}[h!]
  \centering
    \includegraphics[width=0.85\textwidth]{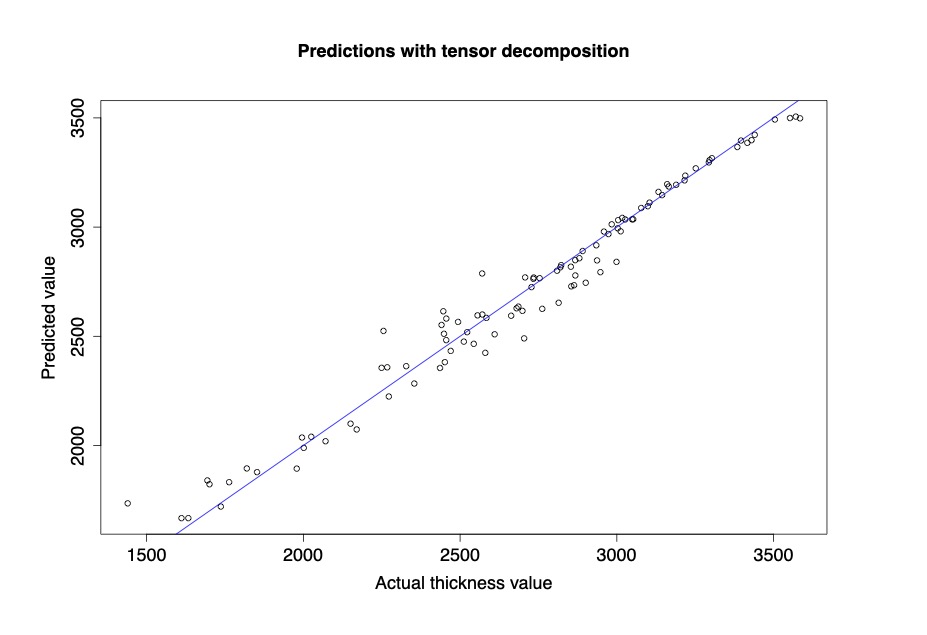}
\caption{Comparison of emulated and actual glacial thickness values, using the tensor-based emulator approach with a mixture of machine learning models.}
\end{figure}

Finally, for an assessment of how the HOSVD tensor-based emulator compares to simply using a standard off-the-shelf supervised learning approach, we compared the absolute relative errors of a GP regression (\textit{kernlab} R package \citep{Karatzoglou:2004aa}) and a random forest (\textit{randomForest} R package \citep{randomForest}). In each of these cases we did not consider a tensor decomposition. Specifically, $10^4$ 5-tuples of $(x,y,t$, period, amplitude) values were sampled with a latin hypercube design to create the training data set.  In all cases, the errors were larger than using the tensor-decomposition; specifically, a random forest produced a mean of 0.0488 absolute relative error (SD: 0.0015), and a GP produced a mean of 0.0387 absolute relative error (SD: 0.000604). Additionally, training data of at least $2 \times 10^4$ training examples resulted in a memory error, suggesting that the tensor emulator method is a more memory efficient procedure.

\subsection{Collective Animal Movement ABM Simulation}\label{sub:animal}

This example considers a collective animal movement ABM based on a simplified version of the model in \citet{COUZIN20021} in which the animals (agents) can exhibit a variety of realistic behaviors such as grouping and repulsion. We first describe the agent-based model for collective behavior, followed by the associated tensor-based emulator, and conclude with an example of using this emulator to learn collective behavior on a real-world data set within a Bayesian inferential paradigm.

\subsubsection{Collective Movement Agent-Based Model Simulator}
Assume we have observed location vectors at time $t=1,\ldots,T$ for the $i$th individual ($i=1,\ldots,N$), denoted by  $\boldsymbol{s}_{i,t} = (x_{i,t},y_{i,t})'$. Our basic model decomposes movement into a unit direction vector and scalar speed for each individual at each discrete time. Specifically, we model the evolution of an agent's location by
$$
\boldsymbol{s}_{i,t} = \boldsymbol{s}_{i,t-1} + \boldsymbol{d}_{i,t}v_{i,t} + \boldsymbol{\eta}_{i,t}, \;\;\; \boldsymbol{\eta}_{i,t} \sim N(0,\sigma^2_\eta \boldsymbol{I}_2),
$$
where $\boldsymbol{d}_{i,t}$ is the unit direction vector, which is a deterministic (black box) function of the most recent past location of all agents and associated parameters (see below), and $v_{i,t}$ is the speed of the agent. In this formulation there is a stochastic component to the location update represented by the $\boldsymbol{\eta}_{i,t}$ error term.  For the simple example presented here we further assume that the speed is constant for all agents across time, $v_{i,t} = v$, and the primary focus is on the unit direction vector, $\boldsymbol{d}_{i,t}$.

\citet{COUZIN20021} notes that an important consideration for animal movement is that the animals do not collide (or, rather, that they have some desired personal space). In addition, animals may favor orientation and attraction to other animals within their perception. That is, there is an intermediate zone beyond the no-collision zone in which animals seek to orient with each other (move in the same direction), and beyond this zone they are attracted to other animals.  In the simple implementation presented here, we consider a no-collision zone and an orientation zone, and assume that individual animals outside of these zones continue moving in the direction they were going at the previous time. 

Let $\delta_{i,j} = ||\boldsymbol{s}_{j,t-1} - \boldsymbol{s}_{i,t-1}||$ be the Euclidean distance between agent $i$ and agent $j$ at time $t-1$. Then, we define the deterministic function for the (un-normalized) direction vector for the $i$th agent at time $t$ by one of the following:
$$
\tilde{\boldsymbol{d}}_{i,t} = \left\{\begin{array}{lr}
- \sum_{j \in {\cal{N}}_{\alpha,-i}} \frac{\boldsymbol{s}_{j,t-1} - \boldsymbol{s}_{i,t-1}}{\delta{i,j}}, &  \mbox{if } \exists \; j \neq i : \delta_{i,j} < \alpha  \\
 \sum_{j \in {\cal N}_{\alpha,\rho_o,i}} \frac{\boldsymbol{d}_{j,t-1}}{||\boldsymbol{d}_{j,t-1}||} &  \alpha \le \delta_{i,j} \le \rho_o \\
 \boldsymbol{d}_{i,t-1} & \delta_{i,j} > \rho_o,
 \end{array} \right.
$$
where $\alpha$ is the no-collision radius, $\rho_o$ is the orientation zone radius,  ${\cal{N}}_{\alpha,-i}$ corresponds to all neighbors of $i$ (not including the $i$th individual) within a distance $\alpha$ of $\boldsymbol{s}_{i,t-1}$, and ${\cal N}_{\alpha,\rho_o,i}$ corresponds to the set of neighbors of $i$ within the distance $(\alpha,\rho_o]$ of the $i$th individual (including the $i$th individual). Thus, the first term corresponds to the no-collision term and applies if {\it any} individual agent is within a distance of $\alpha$ of the $i$th individual.  The second term corresponds to orientation with individuals in a zone between $(\alpha, \rho_o]$ of the $i$th individual (assuming the first condition does not hold). If neither of these conditions hold, the $i$th individual keeps the same direction as before (the third term).  Lastly, we
convert this direction vector to a unit vector
$$
\boldsymbol{d}_{i,t} = \frac{\tilde{\boldsymbol{d}}_{i,t}}{||\tilde{\boldsymbol{d}}_{i,t}||}.
$$

In the simulations considered here, we fixed the no-collision radius ($\alpha$) and the location variance ($\sigma^2_\eta$) and focus on the speed ($v$) and radius of orientation ($\rho_o$), which are the primary drivers of collective behavior. 

\subsubsection{Tensor-Based Emulator}
We train a tensor-based emulator in the following manner:

\begin{enumerate}
\item $2^5$ values of $v$ are sampled uniformly in the range [.1,1], and $2^5$ values of $\rho_o$ are sampled uniformly in the range [5,50]. The parameter $\alpha$ is fixed at .5 and $\sigma^2_\eta$ is fixed at .025. 
\item For each combination of  $v$ and $\rho_o$, and for each time point, the simulator is evaluated for both the $x$ position and $y$ position of the 20 animals. The results of the $x$ positions are stored in a 20 by 101 by 32 by 32 tensor, $\boldsymbol{X}$; likewise, the results of the $y$ positions are stored in a 20 by 101 by 32 by 32 tensor, $\boldsymbol{Y}$. The coordinates of these tensors are individual index, time index, $v$ parameter, and $\rho_o$ parameter.
\item A tensor-emulator is derived using the methodology from Section 3. In particular, $HOSVD(\boldsymbol{X}) = \boldsymbol{Z}\times \boldsymbol{U}_1 \times \boldsymbol{U}_2 \times \boldsymbol{U}_3 \times \boldsymbol{U}_4$, and the predicted $x$ positions for the 20 animals over the 101 times points for parameters $v*$ and $\rho_o*$ is then $\boldsymbol{Z}\times \boldsymbol{U}_1\times \boldsymbol{U}_2\times\boldsymbol{\hat{u}}_3(v*)\times\boldsymbol{\hat{u}}_4(\rho_o*)$, where $\boldsymbol{\hat{u}}_3$(.) and $\boldsymbol{\hat{u}}_4$(.) are functions learned with a machine learning model (see Section 3.3). As in the previous experiment, we considered random forests and GPs, using the default implementations from \citet{randomForest} and \cite{Karatzoglou:2004aa}, respectively. The $y$ positions are emulated with the same procedure, starting with the tensor $\boldsymbol{Y}$.
\end{enumerate}

Two test cases are considered to assess the emulator output for the $x$ and $y$ trajectories. For Case 1 we let $v = .5$ and  $\rho_o = 35$, which forces the animals to move together in a northeast direction given the high degree of collectiveness implied by the larger $\rho_o$ parameter. For test Case 2 we let $v = .5$ and  $\rho_o = 5$, which implies less collectiveness due to the smaller $\rho_o$ parameter, and two groups of animals move in opposite directions. In our assessment of the emulator approaches, we plot the trajectories over time and compare to the original simulated trajectories. Additionally, we compare the simulated and emulated trajectories with measures of collective animal behavior to gain a more nuanced comparison. Specifically, the two quantitative measures used to compare simulated and emulated animal movements through time are the troop spread and troop elongation metrics \citep{strandburg2017habitat}. Troop spread measures how spread apart the group is over time, and troop elongation measures how ellipsoidal the group is over time. 

\begin{figure}[h!]
  \centering
    \includegraphics[width=0.8\textwidth]{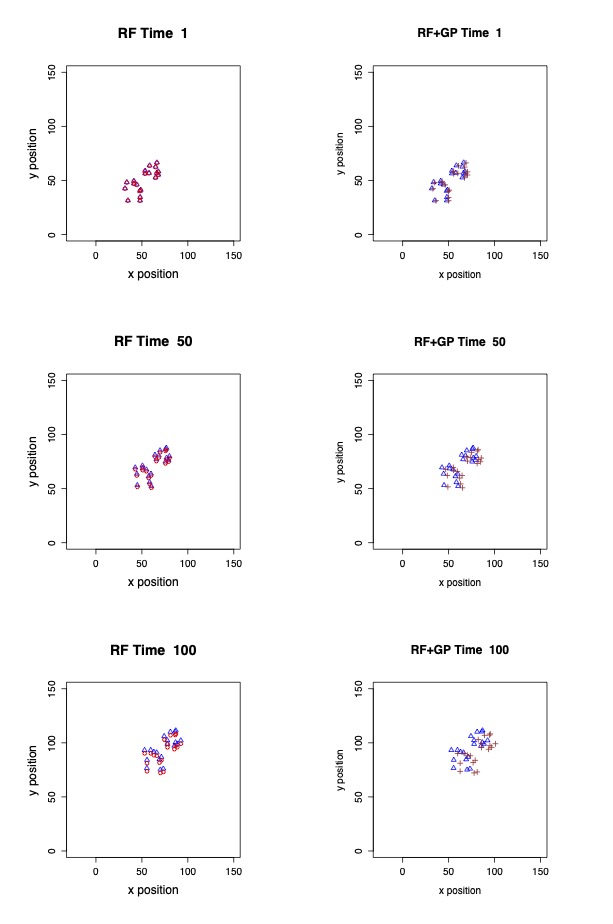}
\caption{Comparison of simulated (triangles) and emulated trajectories for test Case 1, with the pure RF emulator on the left (circles) and the combined model(RF+GP for $x$ positions) emulator on the right (crosses). Both emulator approaches appear to match the simulated positions.}
\label{fig:case1}
\end{figure}

\begin{figure}[h!]
  \centering
    \includegraphics[width=0.85\textwidth]{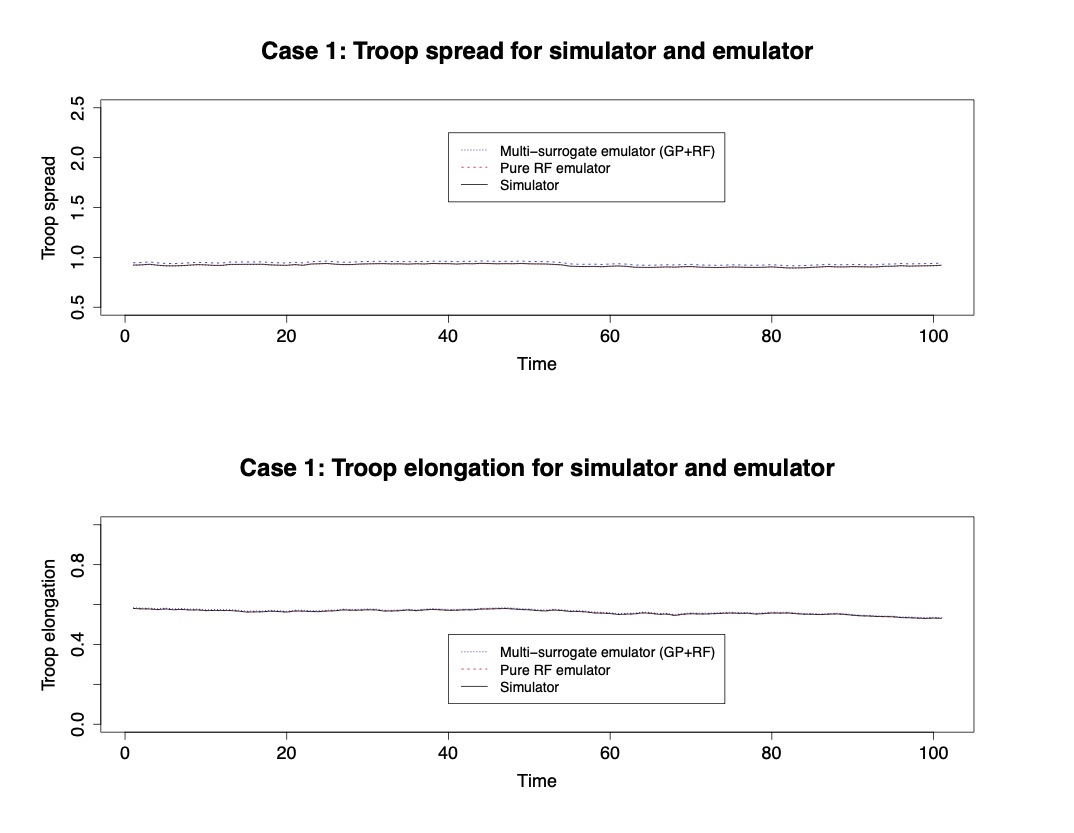}
\caption{Case 1 comparison of troop spread and troop elongation metrics amongst the simulator, pure RF emulator, and multi-model emulator. There is close agreement for both emulators with the simulated trajectories.}
\label{fig:case1_troop}
\end{figure}

Figure \ref{fig:case1} shows a comparison between the simulated and emulated trajectories for test Case 1. We can see that the animals move in a collective group and that the emulated trajectories are quite similar to the true trajectories. This is further demonstrated in Figure  \ref{fig:case1_troop}, which shows a close match for both troop spread and troop elongation between both emulated trajectories and the actual, simulated trajectories. We also compare two emulator approaches with this example. The first is a pure RF for all components, and the second is a multi-model emulator with a RF for the $y$ positions, but a mix of RF for the $\rho_0$ component and GP for the $v$ component of the $x$ positions. That is, the function $\hat{\boldsymbol{u}}_3(v)$ is learned with a GP and $\hat{\boldsymbol{u}}_4(\rho_o)$ is learned with RF regression (for the $x$ positions); we do not restrict these functions to be learned with the same machine learning approach (i.e., RF). Both the location plots in Figure \ref{fig:case1} and comparison of troop spread and troop elongation metrics in Figure \ref{fig:case1_troop} show that there is generally close agreement between both emulator approaches and the simulated trajectories. 

\begin{figure}[h!]
  \centering
    \includegraphics[width=0.75\textwidth]{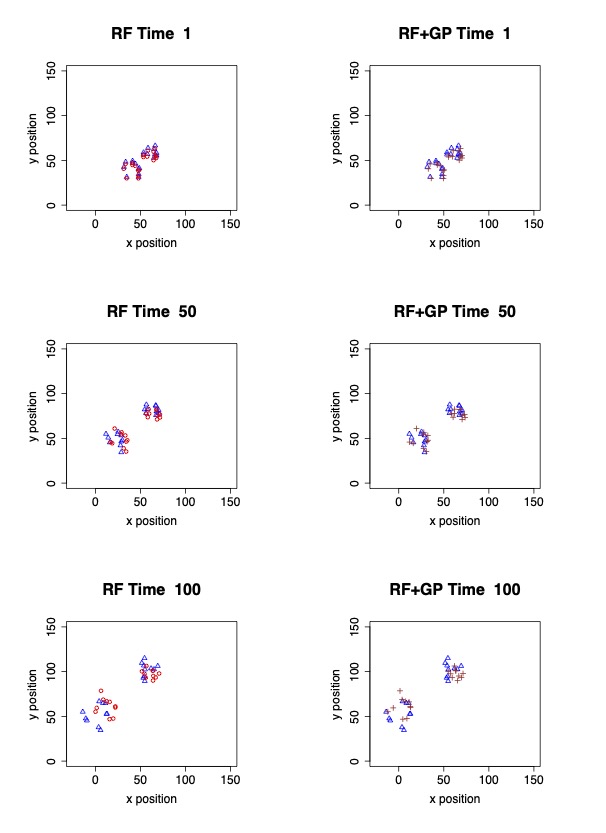}
\caption{Comparison of simulated (triangles) and emulated trajectories for test Case 2, with the pure RF emulator on the left (circles) and the combined model(RF+GP for $x$ positions) emulator on the right (crosses); the tensor-emulator matches the animal trajectories closely. Visually, it is clear that neither emulator matches the simulator as closely as in Case 1, though it is still a good match in terms of capturing the essential collective behavior.}
\label{fig:case2}
\end{figure}

\begin{figure}[h!]
  \centering
    \includegraphics[width=0.85\textwidth]{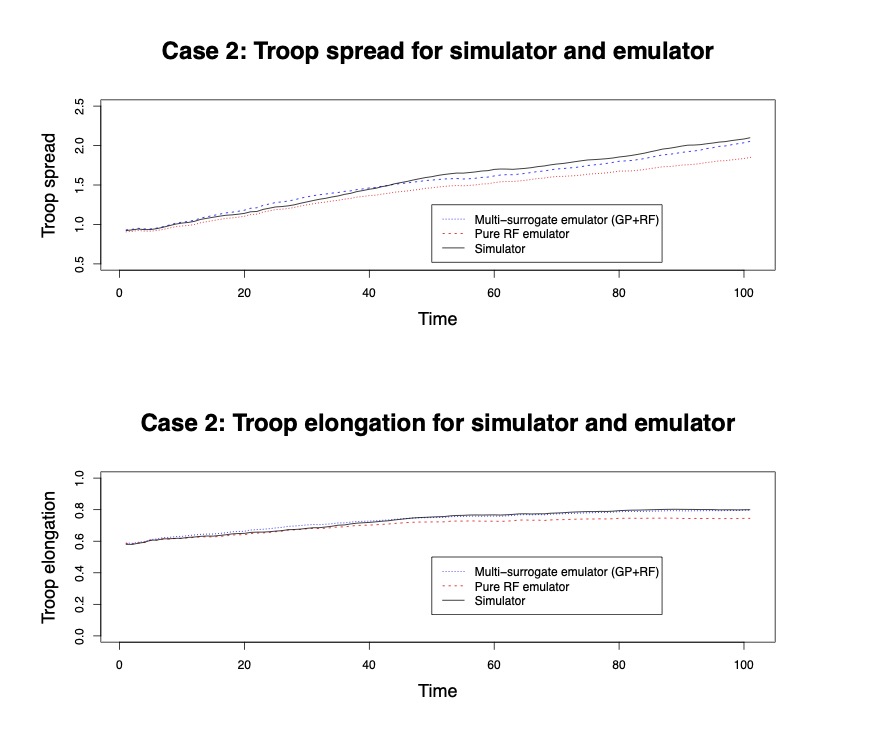}
\caption{Case 2 comparison of troop spread and troop elongation metrics amongst the simulator, pure RF emulator, and multi-model emulator. The troop spread and troop elongation profiles are best represented by the multi-model emulator.}
\label{fig:case2_troop}
\end{figure}

In Case 2, simple visual inspection of both emulator trajectories in Figure \ref{fig:case2} makes it difficult to conclude which emulator (i.e, pure RF or combined RF/GP) emulates the true trajectories closest. However, comparisons of troop spread and troop elongation in Figure \ref{fig:case2_troop} generally show that the combined RF/GP emulator better captures the simulator than the pure RF surrogate emulator, further illustrating the potential utility of a combined RF/GP approach over using a single machine learning model. 

Additionally, we use Case 1 and Case 2 to make a comparison to a widely-used GP emulator, \texttt{ppgasp} from the \texttt{RobustGaSP} \texttt{R} package \citep{gu2018robustgasp}. Specifically, for both synthetic test cases we calculate the mean distances between the emulated trajectories and the simulated trajectories for all of the animals and over all time points, and refer to this as mean error. It appears that the tensor-emulator is more accurate in both Case 1 and Case 2, though with most improvement in Case 2. However, ppgasp has a much faster emulator evaluation time in both test cases. These results are summarized in Table 1. A faster compute time for emulator evaluation could be achieved with a lower rank HOSVD decomposition, though that would likely come with some accuracy trade-off. Additionally, an important consideration is that output is restricted to be on a temporal grid for all of the animals for the sake of comparison, though in contrast to \texttt{ppgasp}, the HOSVD emulator handles the scenario where time is not assumed to be on a discrete grid. 

\begin{table}[]
\centering
\begin{tabular}{|c|c|c|c|}
\hline
Case & Emulator       & Mean Error & Run-time   \\ \hline
1    & HOSVD          & 1.678      & 0.245      \\ \hline
1    & ppgasp         & 2.762      & 0.0072     \\ \hline
2    & HOSVD          & 5.969      & 0.2423     \\ \hline
2    & ppgasp         & 22.33      & 0.0077     \\ \hline
     & \textbf{Units} & \textbf{m} & \textbf{s} \\ \hline
\end{tabular}
\caption{Comparison of run-time and accuracy of the ppgasp and HOSVD emulators on two synthetic ABM test cases. Run-times are listed for evaluation of emulator.}
\label{tab:table1}
\end{table}

\subsubsection{Bayesian Learning with the Tensor-Based Emulator}
In order to test the emulator on real data in a Bayesian inferential setting, we used the emulator version of our simplified animal movement ABM on guppy ({\it Poecilia reticulata}) data; specifically, experiment fm7 from \citet{bode2012distinguishing}. These data come from an experiment using a captive population of guppies. In this experiment, groups of ten guppies of the same sex were filmed from above in a square tank in which one corner contained gravel and shade, which is presumed to be attractive to the guppies because it provides shelter.  The guppies were released in the tank in the opposite (lower-right) corner. The data consist of movement trajectories truncated to the time points when all individuals were moving until one guppy reached the shaded target area.

In particular, we assume the following data model for the guppy trajectories:
\begin{eqnarray*}
O_{itx} &=& \hat{f}_x(i,t,v,\rho_0)+\epsilon_{itx}, \;\; \epsilon_{itx} \sim iid \; N(0,\sigma^2_x) \\
O_{ity} &=& \hat{f}_y(i,t,v,\rho_0)+\epsilon_{ity}, \;\; \epsilon_{ity} \sim iid \; N(0,\sigma^2_y), \\
\end{eqnarray*}
where $O_{itx}$ and $O_{ity}$ are observed $x$ and $y$ locations for the $i$th guppy at time $t$, respectively, and  $\hat{f}_x$ and $\hat{f}_y$ are tensor emulators constructed in the same manner as the previous section (trained with runs from the animal movement simulator described previously).  Additionally, we assume the following prior distributions on parameters: $\rho_o \sim \mbox{Uniform}(5,60)$, $v \sim \mbox{Uniform}(1,5)$, $\sigma^2_x \sim \mbox{InverseGamma(shape=3, scale=4)}$, and $\sigma^2_y \sim \mbox{InverseGamma(shape=3, scale=4)}$. 

\begin{figure}[h!]
  \centering
    \includegraphics[width=0.85\textwidth]{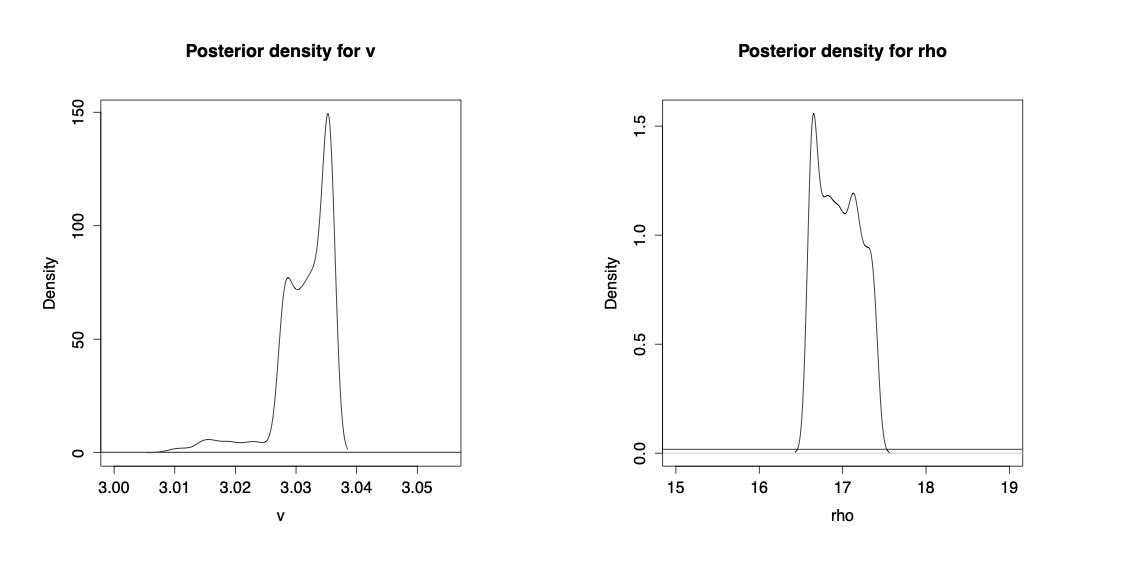}
\caption{Posterior densities for both $\rho_o$ and $v$ are illustrated, which are inferred using guppy data (fm7) from \citet{bode2012distinguishing}. The posterior for both parameters concentrates in a much more narrow range than the uniform prior, suggesting Bayesian learning.}
\label{fig:gup_post}
\end{figure}

\begin{figure}[h!]
  \centering
    \includegraphics[width=0.99\textwidth]{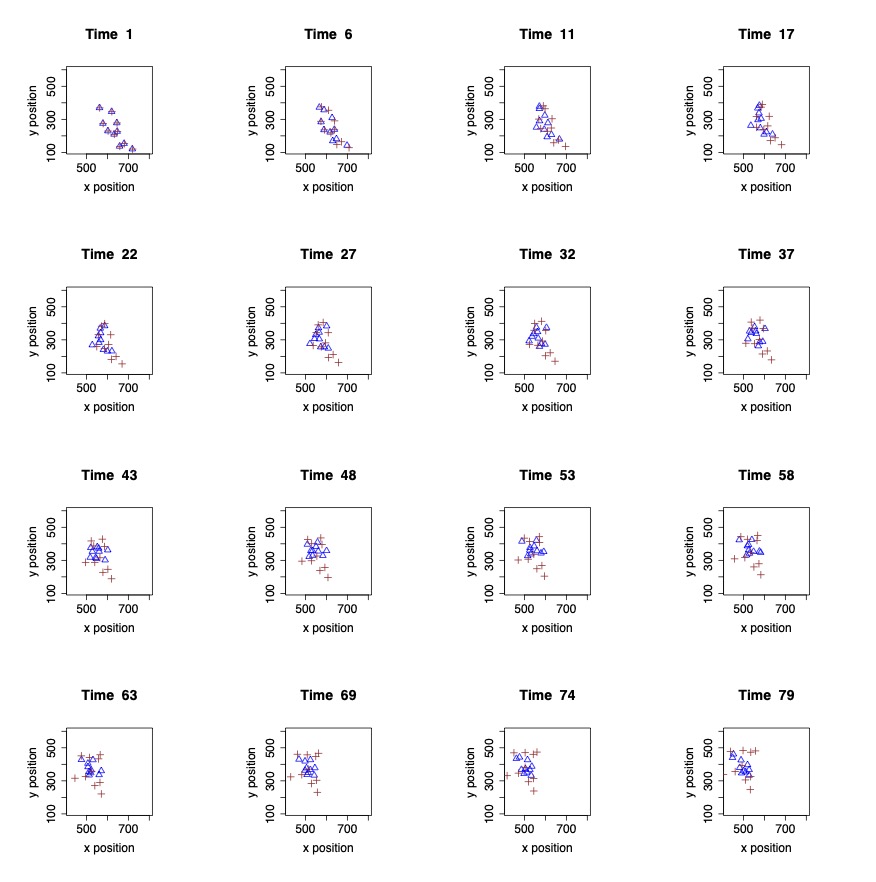}
\caption{Simulated trajectories for guppy data (fm7 from \citet{bode2012distinguishing}) using posterior means of $\rho_0$ and $v$ compared to the real data, which indicate a similar pattern in collective behavior. The triangles are the actual trajectories, and the crosses are the simulated trajectories using the posterior means of parameters.}
\label{fig:gup_trajectory}
\end{figure}

We implement a Gibbs sampler for sampling from the posterior using inverse-gamma draws (due to conjugacy) for $\sigma^2_x$ and $\sigma^2_y$ and grid sampling steps for $\rho_o$ and $v$. Traceplots, given in the supplementary materials, show good mixing, and the resultant posterior distributions for $\rho_o$ and $v$ are illustrated in Figure \ref{fig:gup_post}. The relatively narrow regions where mass concentrates (in comparison to the uniform priors) suggests that there is definite Bayesian learning of these parameters, though since this is real data, there are no ground-truth parameters to compare to. However, the posterior mean of the $\rho_o$ orientation distance parameter suggests that there is a substantial amount of collective behavior by these guppies in this experiment. A plot of the simulated trajectory in comparison to the emulated trajectory (using the posterior mean of $\rho_o$ and $v$) is included in Figure \ref{fig:gup_trajectory}, and indicates overall agreement in the patterns of collective behavior.  

\section{Conclusion}
The objective of this paper has been to introduce a method to construct an emulator for a complex spatio-temporal function (e.g., computer simulator) using tensor decomposition with potentially different surrogate models for each tensor dimension. In particular, the method is essentially a variant of a SVD-based emulator approach \citep{higdon2008computer} that uses the HOSVD intead of an SVD. The distinguishing features of the tensor-based method for emulation are:
\begin{itemize}
\item The method can predict the function output at spatial locations and time points not considered in the training sample, in contrast to many existing emulator methods.
\item A variety of supervised learning approaches (e.g., GP regression and random forests) can be combined for the spatial, temporal, and parameter components. This allows greater flexibility and the ability to combine the strengths (and disadvantages) of different machine learning approaches in constructing an emulator, i.e., a ``mix and match'' approach.
\item The method can provide efficient emulation of the spatio-temporal locations of multiple individuals in an ABM.
\item By using a reduced rank tensor factorization with the truncated HOSVD \citep{de2000multilinear}, computational savings are achieved.
\end{itemize}
The method has been demonstrated via application in two scenarios: a periodic solution to a shallow ice approximation partial differential equation from glaciology and a collective animal movement ABM simulator. In both scenarios, combining supervised learning approaches yields an accurate emulator, demonstrating the advantage of model flexibility when different machine learning approaches are combined. It is conceivable the same results will hold for more complex supervised learning methods such as deep learning; in fact, an advantage of the proposed approach is that it can handle the use of a variety of machine learning models, and so is adaptable to the rapid advances being made in the field. While we conducted some preliminary experiments using the \texttt{neuralnetwork} R package with default options, the results were generally poorer than the \texttt{randomForest} and \texttt{kernlab} defaults for random forests and GP regression; presumably more advanced neural network architectures will show neural networks to be much more competitive. An additional strength of the methodology presented is that it has been implemented with R packages available on the Comprehensive R Archive Network (CRAN), so a tensor-based emulator can be directly included in an application without much difficulty, for instance in a spatio-temporal Bayesian hierarchical model as in \cite{gopalan2019hierarchical}. This was illustrated in the collective animal movement example with an experiment that documented the movement of guppies in a tank.  The Bayesian implementation with the tensor-based emulator suggested that the guppies exhibit a high-degree of collectiveness in this experiment. While we have not covered emulator uncertainty in this treatment, we believe the literature on model discrepancy \citep{kennedy2001bayesian, discrepancy, gopalan2019hierarchical} can be used to construct hierarchical models involving an emulator that accounts for emulator uncertainty and inaccuracy. 
\bibliographystyle{jasa3}
\bibliography{references.bib}

\begin{thebibliography}{36}
\newcommand{\enquote}[1]{``#1''}
\expandafter\ifx\csname natexlab\endcsname\relax\def\natexlab#1{#1}\fi
\expandafter\ifx\csname url\endcsname\relax
  \def\url#1{{\tt #1}}\fi
\expandafter\ifx\csname urlprefix\endcsname\relax\def\urlprefix{URL }\fi

\bibitem[\protect\citeauthoryear{Aðalgeirsd{\'o}ttir}{Aðalgeirsd{\'o}ttir}{2003}]{Tolly}
Aðalgeirsd{\'o}ttir, G. (2003), \enquote{Flow dynamics of {Vatnaj{\"o}kull}
  ice cap, {Iceland},} {\em Mitteilungen der Versuchsanstalt fur Wasserbau,
  Hydrologie und Glaziologie an der Eidgenossischen Technischen Hochschule
  Zurich\/}.

\bibitem[\protect\citeauthoryear{Berliner}{Berliner}{1996}]{10.1007/978-94-011-5430-7_3}
Berliner, L.~M. (1996), \enquote{Hierarchical {Bayesian} Time Series Models,}
  in Hanson, K.~M. and Silver, R.~N. (editors), {\em Maximum Entropy and
  Bayesian Methods\/}, Dordrecht: Springer Netherlands.

\bibitem[\protect\citeauthoryear{Berliner}{Berliner}{2003}]{Berliner}
--- (2003), \enquote{Physical-statistical modeling in geophysics,} {\em Journal
  of Geophysical Research: Atmospheres\/}, 108, n/a--n/a,
  \urlprefix\url{http://dx.doi.org/10.1029/2002JD002865}. 8776.

\bibitem[\protect\citeauthoryear{Bode, Franks, Wood, Piercy, Croft, and
  Codling}{Bode et~al.}{2012}]{bode2012distinguishing}
Bode, N.~W., Franks, D.~W., Wood, A.~J., Piercy, J.~J., Croft, D.~P., and
  Codling, E.~A. (2012), \enquote{Distinguishing social from nonsocial
  navigation in moving animal groups,} {\em The American Naturalist\/}, 179,
  621--632.

\bibitem[\protect\citeauthoryear{Brynjarsd{\'o}ttir and
  O'Hagan}{Brynjarsd{\'o}ttir and O'Hagan}{2014}]{discrepancy}
Brynjarsd{\'o}ttir, J. and O'Hagan, A. (2014), \enquote{Learning about physical
  parameters: the importance of model discrepancy,} {\em Inverse Problems\/},
  30, 114007, \urlprefix\url{http://stacks.iop.org/0266-5611/30/i=11/a=114007}.

\bibitem[\protect\citeauthoryear{{Bueler}, {Lingle}, {Kallen-Brown}, {Covey},
  and {Bowman}}{{Bueler} et~al.}{2005}]{Bueler}
{Bueler}, E., {Lingle}, C.~S., {Kallen-Brown}, J.~A., {Covey}, D.~N., and
  {Bowman}, L.~N. (2005), \enquote{{Exact solutions and verification of
  numerical models for isothermal ice sheets},} {\em Journal of Glaciology\/},
  51, 291--306.

\bibitem[\protect\citeauthoryear{Chevreuil, Lebrun, Nouy, and Rai}{Chevreuil
  et~al.}{2015}]{chevreuil2015least}
Chevreuil, M., Lebrun, R., Nouy, A., and Rai, P. (2015), \enquote{A
  least-squares method for sparse low rank approximation of multivariate
  functions,} {\em SIAM/ASA Journal on Uncertainty Quantification\/}, 3,
  897--921.

\bibitem[\protect\citeauthoryear{Couzin, Krause, James, Ruxton, and
  Franks}{Couzin et~al.}{2002}]{COUZIN20021}
Couzin, I.~D., Krause, J., James, R., Ruxton, G.~D., and Franks, N.~R. (2002),
  \enquote{Collective Memory and Spatial Sorting in Animal Groups,} {\em
  Journal of Theoretical Biology\/}, 218, 1 -- 11,
  \urlprefix\url{http://www.sciencedirect.com/science/article/pii/S0022519302930651}.

\bibitem[\protect\citeauthoryear{De~Lathauwer, De~Moor, and
  Vandewalle}{De~Lathauwer et~al.}{2000}]{de2000multilinear}
De~Lathauwer, L., De~Moor, B., and Vandewalle, J. (2000), \enquote{A
  multilinear singular value decomposition,} {\em SIAM journal on Matrix
  Analysis and Applications\/}, 21, 1253--1278.

\bibitem[\protect\citeauthoryear{Fadikar, Higdon, Chen, Lewis, Venkatramanan,
  and Marathe}{Fadikar et~al.}{2018}]{fadikar2018calibrating}
Fadikar, A., Higdon, D., Chen, J., Lewis, B., Venkatramanan, S., and Marathe,
  M. (2018), \enquote{Calibrating a stochastic, agent-based model using
  quantile-based emulation,} {\em SIAM/ASA Journal on Uncertainty
  Quantification\/}, 6, 1685--1706.

\bibitem[\protect\citeauthoryear{Flowers, Marshall, Bj{\"o}rnsson, and
  Clarke}{Flowers et~al.}{2005}]{flowers2005sensitivity}
Flowers, G.~E., Marshall, S.~J., Bj{\"o}rnsson, H., and Clarke, G.~K. (2005),
  \enquote{{Sensitivity of {Vatnaj{\"o}kull} ice cap hydrology and dynamics to
  climate warming over the next 2 centuries},} {\em Journal of Geophysical
  Research: Earth Surface\/}, 110.

\bibitem[\protect\citeauthoryear{Gopalan, Hrafnkelsson, Wikle, Rue,
  A{\dh}algeirsd{\'o}ttir, Jarosch, and P{\'a}lsson}{Gopalan
  et~al.}{2019}]{gopalan2019hierarchical}
Gopalan, G., Hrafnkelsson, B., Wikle, C.~K., Rue, H., A{\dh}algeirsd{\'o}ttir,
  G., Jarosch, A.~H., and P{\'a}lsson, F. (2019), \enquote{{A Hierarchical
  Spatiotemporal Statistical Model Motivated by Glaciology},} {\em Journal of
  Agricultural, Biological and Environmental Statistics\/}, 24, 669--692.

\bibitem[\protect\citeauthoryear{Gramacy}{Gramacy}{2020}]{gramacy2020surrogates}
Gramacy, R.~B. (2020), {\em Surrogates: {G}aussian Process Modeling, Design and
  Optimization for the Applied Sciences\/}, Boca Raton, Florida: Chapman
  Hall/CRC. \url{http://bobby.gramacy.com/surrogates/}.

\bibitem[\protect\citeauthoryear{Gu, Palomo, and Berger}{Gu
  et~al.}{2018{\natexlab{a}}}]{gu2018robustgasp}
Gu, M., Palomo, J., and Berger, J.~O. (2018{\natexlab{a}}),
  \enquote{RobustGaSP: Robust Gaussian stochastic process emulation in R,} {\em
  arXiv preprint arXiv:1801.01874\/}.

\bibitem[\protect\citeauthoryear{Gu, Wang, and Berger}{Gu
  et~al.}{2018{\natexlab{b}}}]{gu2018}
Gu, M., Wang, X., and Berger, J.~O. (2018{\natexlab{b}}), \enquote{{Robust
  Gaussian stochastic process emulation},} {\em Annals of Statistics\/}, 46,
  3038--3066, \urlprefix\url{https://doi.org/10.1214/17-AOS1648}.

\bibitem[\protect\citeauthoryear{{Guan}, {Haran}, and {Pollard}}{{Guan}
  et~al.}{2016}]{2016arXiv161201454G}
{Guan}, Y., {Haran}, M., and {Pollard}, D. (2016), \enquote{{Inferring Ice
  Thickness from a Glacier Dynamics Model and Multiple Surface Datasets},} {\em
  ArXiv e-prints\/}.

\bibitem[\protect\citeauthoryear{Hastie, Tibshirani, and Friedman}{Hastie
  et~al.}{2009}]{hastie2009elements}
Hastie, T., Tibshirani, R., and Friedman, J. (2009), {\em The Elements of
  Statistical Learning: Data Mining, Inference, and Prediction\/}, Springer
  Science \& Business Media.

\bibitem[\protect\citeauthoryear{Higdon, Gattiker, Williams, and
  Rightley}{Higdon et~al.}{2008}]{higdon2008computer}
Higdon, D., Gattiker, J., Williams, B., and Rightley, M. (2008),
  \enquote{Computer model calibration using high-dimensional output,} {\em
  Journal of the American Statistical Association\/}, 103, 570--583.

\bibitem[\protect\citeauthoryear{Hooten, Leeds, Fiechter, and Wikle}{Hooten
  et~al.}{2011}]{Hooten2011}
Hooten, M.~B., Leeds, W.~B., Fiechter, J., and Wikle, C.~K. (2011),
  \enquote{Assessing First-Order Emulator Inference for Physical Parameters in
  Nonlinear Mechanistic Models,} {\em Journal of Agricultural, Biological, and
  Environmental Statistics\/}, 16, 475--494,
  \urlprefix\url{https://doi.org/10.1007/s13253-011-0073-7}.

\bibitem[\protect\citeauthoryear{Jarosch, Schoof, and Anslow}{Jarosch
  et~al.}{2013}]{alex}
Jarosch, A.~H., Schoof, C.~G., and Anslow, F.~S. (2013), \enquote{{Restoring
  mass conservation to shallow ice flow models over complex terrain},} {\em The
  Cryosphere\/}, 7, 229--240,
  \urlprefix\url{https://www.the-cryosphere.net/7/229/2013/}.

\bibitem[\protect\citeauthoryear{Karatzoglou, Smola, Hornik, and
  Zeileis}{Karatzoglou et~al.}{2004}]{Karatzoglou:2004aa}
Karatzoglou, A., Smola, A., Hornik, K., and Zeileis, A. (2004),
  \enquote{kernlab -- An {S4} Package for Kernel Methods in {R},} {\em Journal
  of Statistical Software\/}, 11, 1--20,
  \urlprefix\url{http://www.jstatsoft.org/v11/i09/}.

\bibitem[\protect\citeauthoryear{Kasim, Watson-Parris, Deaconu, Oliver,
  Hatfield, Froula, Gregori, Jarvis, Khatiwala, Korenaga et~al.}{Kasim
  et~al.}{2020}]{kasim2020up}
Kasim, M., Watson-Parris, D., Deaconu, L., Oliver, S., Hatfield, P., Froula,
  D., Gregori, G., Jarvis, M., Khatiwala, S., Korenaga, J., et~al. (2020),
  \enquote{Up to two billion times acceleration of scientific simulations with
  deep neural architecture search,} {\em arXiv preprint arXiv:2001.08055\/}.

\bibitem[\protect\citeauthoryear{Kennedy and O'Hagan}{Kennedy and
  O'Hagan}{2001}]{kennedy2001bayesian}
Kennedy, M.~C. and O'Hagan, A. (2001), \enquote{{Bayesian} calibration of
  computer models,} {\em Journal of the Royal Statistical Society: Series B
  (Statistical Methodology)\/}, 63, 425--464.

\bibitem[\protect\citeauthoryear{Kolda and Bader}{Kolda and
  Bader}{2009}]{kolda2009tensor}
Kolda, T.~G. and Bader, B.~W. (2009), \enquote{Tensor decompositions and
  applications,} {\em SIAM review\/}, 51, 455--500.

\bibitem[\protect\citeauthoryear{Konakli and Sudret}{Konakli and
  Sudret}{2016}]{KONAKLI20161144}
Konakli, K. and Sudret, B. (2016), \enquote{Polynomial meta-models with
  canonical low-rank approximations: Numerical insights and comparison to
  sparse polynomial chaos expansions,} {\em Journal of Computational
  Physics\/}, 321, 1144 -- 1169,
  \urlprefix\url{http://www.sciencedirect.com/science/article/pii/S0021999116302303}.

\bibitem[\protect\citeauthoryear{Kuhnert}{Kuhnert}{2014}]{kuhnert2014physical}
Kuhnert, P.~M. (2014), \enquote{Physical-Statistical Modeling,} {\em Wiley
  StatsRef: Statistics Reference Online\/}, 1--5.

\bibitem[\protect\citeauthoryear{Leeds, Wikle, and Fiechter}{Leeds
  et~al.}{2014}]{leeds2014emulator}
Leeds, W.~B., Wikle, C.~K., and Fiechter, J. (2014), \enquote{Emulator-assisted
  reduced-rank ecological data assimilation for nonlinear multivariate
  dynamical spatio-temporal processes,} {\em Statistical Methodology\/}, 17,
  126--138.

\bibitem[\protect\citeauthoryear{Li, Bien, and Wells}{Li
  et~al.}{2018}]{JSSv087i10}
Li, J., Bien, J., and Wells, M. (2018), \enquote{{rTensor: An R Package for
  Multidimensional Array (Tensor) Unfolding, Multiplication, and
  Decomposition},} {\em Journal of Statistical Software, Articles\/}, 87,
  1--31, \urlprefix\url{https://www.jstatsoft.org/v087/i10}.

\bibitem[\protect\citeauthoryear{Liaw and Wiener}{Liaw and
  Wiener}{2002}]{randomForest}
Liaw, A. and Wiener, M. (2002), \enquote{Classification and Regression by
  randomForest,} {\em R News\/}, 2, 18--22,
  \urlprefix\url{https://CRAN.R-project.org/doc/Rnews/}.

\bibitem[\protect\citeauthoryear{Pratola and Chkrebtii}{Pratola and
  Chkrebtii}{2018}]{pratola2018bayesian}
Pratola, M.~T. and Chkrebtii, O. (2018), \enquote{Bayesian calibration of
  multistate stochastic simulators,} {\em Statistica Sinica\/}, 693--719.

\bibitem[\protect\citeauthoryear{Salter, Williamson, Scinocca, and
  Kharin}{Salter et~al.}{2019}]{doi:10.1080/01621459.2018.1514306}
Salter, J.~M., Williamson, D.~B., Scinocca, J., and Kharin, V. (2019),
  \enquote{{Uncertainty Quantification for Computer Models With Spatial Output
  Using Calibration-Optimal Bases},} {\em Journal of the American Statistical
  Association\/}, 0, 1--24,
  \urlprefix\url{https://doi.org/10.1080/01621459.2018.1514306}.

\bibitem[\protect\citeauthoryear{Sargsyan}{Sargsyan}{2016}]{Sargsyan2016}
Sargsyan, K. (2016), {\em {Surrogate Models for Uncertainty Propagation and
  Sensitivity Analysis}\/}, Cham: Springer International Publishing, 1--26.

\bibitem[\protect\citeauthoryear{Strandburg-Peshkin, Farine, Crofoot, and
  Couzin}{Strandburg-Peshkin et~al.}{2017}]{strandburg2017habitat}
Strandburg-Peshkin, A., Farine, D.~R., Crofoot, M.~C., and Couzin, I.~D.
  (2017), \enquote{Habitat and social factors shape individual decisions and
  emergent group structure during baboon collective movement,} {\em Elife\/},
  6, e19505.

\bibitem[\protect\citeauthoryear{Tripathy and Bilionis}{Tripathy and
  Bilionis}{2018}]{tripathy2018deep}
Tripathy, R.~K. and Bilionis, I. (2018), \enquote{Deep UQ: Learning deep neural
  network surrogate models for high dimensional uncertainty quantification,}
  {\em Journal of Computational Physics\/}, 375, 565--588.

\bibitem[\protect\citeauthoryear{Werder, Huss, Paul, Dehecq, and
  Farinotti}{Werder et~al.}{2020}]{werder_huss_paul_dehecq_farinotti_2020}
Werder, M.~A., Huss, M., Paul, F., Dehecq, A., and Farinotti, D. (2020),
  \enquote{{A Bayesian ice thickness estimation model for large-scale
  applications},} {\em Journal of Glaciology\/}, 66, 137–152.

\bibitem[\protect\citeauthoryear{Wikle and Hooten}{Wikle and
  Hooten}{2010}]{wikle2010general}
Wikle, C.~K. and Hooten, M.~B. (2010), \enquote{A general science-based
  framework for dynamical spatio-temporal models,} {\em Test\/}, 19, 417--451.

\end{thebibliography}

\end{document}